\documentclass{PoS}

\title{Accretion onto deformed black holes via pseudo--Newtonian potentials}

\ShortTitle{Pseudo--Newtonian accretion}

\author{\speaker{Anslyn J. John} 
\\
        Department of Mathematics, Rhodes University, Grahamstown 6139, South Africa\\
        E-mail: \email{a.john@ru.ac.za}}

\author{Chris Z. Stevens \\
        Department of Mathematics, Rhodes University, Grahamstown 6139, South Africa\\
        E-mail: \email{c.stevens@ru.ac.za}}

\abstract{The Johannsen--Psaltis spacetime describes a rotating black hole with parametric deviations from the Kerr metric. By construction this spacetime explicitly violates the no-hair theorems. Rotating black hole solutions in any modified theory of gravity could be written in terms of the Johannsen--Psaltis metric. We examined the accretion of gas onto a black hole described by the static limit of this spacetime. We employed a potential that generalises the Paczy{\'n}ski--Wiita potential to the static Johannsen--Psaltis metric. Our analysis utilised a recent pseudo--Newtonian formulation of the dynamics around arbitrary static, spherically symmetric spacetimes. We found that positive (negative) values of the scalar hair parameter, $\epsilon_{3}$ increased (decreased) the accretion rate. This framework can be extended to incorporate various astrophysical phenomena like radiative processes, viscous dissipation and magnetic fields.}

\FullConference{High Energy Astrophysics in Southern Africa - HEASA2018\\
		1-3 August, 2018\\
		Parys, Free State, South Africa}

\begin{document}

\section{Introduction}

General relativity has been successfully tested in weak gravitational fields on scales of the size of the solar system \cite{Will2014}. The first strong field tests have become feasible following the discovery of gravitational radiation from merging binary black holes \cite{abbott2016observation}. Black holes in general relativity are described by the Schwarzschild solution if they are static, or by the Kerr solution if they are rotating. The Kerr metric is the only (uncharged) stationary, axisymmetric, asymptotically flat vacuum spacetime in general relativity possessing an event horizon. This is a consequence of the no--hair theorem \cite{nohair1, nohair2, nohair3}. 

Violations of the no--hair theorem necessarily imply the breakdown of general relativity. Modified theories of gravity admit black hole solutions that possess scalar hair. Given the advent of gravitational wave astronomy it has become feasible to search for black holes with scalar hair with current and future generations of detectors \cite{berti2015testing,broderick2014testing}. Gravitational lensing by black holes with scalar hair would produce shadows (images of the black hole photon sphere) that are distinct from those of black holes in general relativity. Different shapes, sizes and multiple images of the shadow can appear \cite{cunha2015shadows}. The first image of the shadow of a black hole was observed by the Event Horizon Telescope in 2019 \cite{akiyama2019firstI,akiyama2019firstII,akiyama2019firstIII,akiyama2019firstIV,akiyama2019firstV,akiyama2019firstVI} showing no significant deviations from the shadow shapes predicted by general relativity. This puts the scientific community in a position to eliminate some alternative theories of gravity.

A plethora of modified theories of gravity has been proposed \cite{moffat2006scalar, teves, nojiri}. Determining the precise form of their possible black hole solutions as well as their observational signatures is a formidable task. A complementary, bottom--up approach is to construct generic black hole spacetimes without appealing to any specific theory of gravity and determine their observational features. By constraining the magnitude of inferred deviations from the Kerr metric, one can determine limits on modified theories of gravity. 

Generic black hole spacetimes are deformed in the sense that they parametrically deviate from the Kerr spacetime. The Johannsen--Psaltis (JP) metric \cite{johannsen2011metric} is an example of such a spacetime. It facilitates model--independent tests of gravity. The JP metric does not arise from a particular theory of gravity and violates the no--hair theorem by construction. In contrast to a number of alternative deformed black hole spacetimes \cite{manko1992generalizations}, the JP metric remains free of various pathologies like singularities and closed timelike curves outside the event horizon \cite{johannsen2011metric}. 

In this paper we study accretion onto a JP black hole using a pseudo--Newtonian gravitational potential. Models of accreting systems become extremely sophisticated due to the presence of features like viscosity, magnetic fields, turbulence and radiative processes. These models become even more complex when one incorporates full general relativity. Paczy{\'n}ski and Wiita \cite{paczynsky1980} introduced a potential that effectively mimics many of the gravitational features of the Schwarzschild black hole. This potential has been commonly used to successfully model accretion onto black holes \cite{abramowicz2009paczynski}. The equations of motion closely resemble the corresponding ones in the Newtonian problem and are significantly easier to analyse than those that arise in a fully general relativistic treatment.

Recently a number of authors \cite{faraoni2016paczynski,tejeda2014} devised algorithms to generate pseudo--Newtonian potentials that approximate the behaviour of black holes in any static, spherically symmetric spacetime. These potentials are generalizations of the Paczy{\'n}ski--Wiita potential, which approximates the Schwarzschild metric. Whilst the Paczy{\'n}ski--Wiita potential was originally derived in an ad hoc fashion \cite{paczynsky1980}, its specific form can be justified quite rigorously \cite{abramowicz2009paczynski}. The main idea is to obtain the geodesic equations describing the motion a massive test particle around a Schwarzschild black hole and compare them to the equivalent orbital equations in Newtonian gravity. One then identifies the term in the (fully relativistic) geodesic equation that corresponds to the potential in the Newtonian equation of motion. This term, which is defined to be the effective potential, turns out to be the Paczy{\'n}ski--Wiita potential \cite{faraoni2016paczynski}. 

The Paczy{\'n}ski--Wiita potential predicts precisely the same values for the innermost stable circular orbit (ISCO), the marginally bound orbit ($r_{mb}$), and the Keplerian angular momentum, $L(r)$, as those obtained from a fully general relativistic analysis of the Schwarzschild geodesic equations \cite{abramowicz2009paczynski}. Also the fixed points of the associated dynamical systems are identical \cite{abramowicz2009paczynski}. 

The Paczy{\'n}ski--Wiita potential, $\Phi_{PW}$, is not the only pseudo--Newtonian potential devised, but it is by far the most popular. Other proposals tend to be obtained by fitting formulae. The $\Phi_{PW}$ potential has some limitations. It cannot properly account for the dynamics around rotating black holes as it fails to predict the Lense--Thirring effect, nor can it accurately describe self--gravitating systems \cite{abramowicz2009paczynski}.

In this paper we derive a pseudo--Newtonian potential for the static limit of the JP spacetime and solve the associated accretion problem. This is an important first approximation to more complex accretion problems. Many of the features of this idealized scenario survive (e.g. the existence of a transonic solution) in the more complicated accretion disk problem. We use geometric units throughout this paper where $c=1=G$.

\section{Pseudo--Newtonian potentials}

The static, spherically symmetric form of the JP metric is
\begin{eqnarray}\label{eq:SSJPmetric}
	ds^2= &- [1+h(r)]\Big{(}1 - \frac{2m}{r}\Big{)}dt^2 
		  + [1+h(r)]\Big{(}1 - \frac{2m}{r}\Big{)}^{-1}dr^2 \nonumber \\
		  &+ r^2\Big{(}d\theta^2 + \sin^2\theta d\phi^2\Big{)},
\end{eqnarray}
where the parametric deviation is given by
\begin{equation}\label{eq:SSdeviationh}
	h(r) = \sum_{k=0}^{\infty}\Big{(}\epsilon_{2k} 
				   + \epsilon_{2k+1}\frac{m}{r}\Big{)}\Big{(}\frac{m}{r}\Big{)}^{2k}.
\end{equation}
The line element of the Schwarzschild black hole is recovered when $\epsilon_{k} = 0$. There are a number of observational constraints on the magnitude of the dimensionless $\epsilon_{k}$ parameters \cite{johannsen2011metric}. Requiring the spacetime to be stationary and asymptotically flat forces $\epsilon_{0} = \epsilon_{1} = 0$. 
The Lunar Laser Ranging experiment \cite{Will2014} sets an upper bound of $|\epsilon_{2}| \leq 4.6 \times 10^{-4}$. The first unconstrained parameter in the expansion of $h(r)$ is $\epsilon_{3}$. The scalar hair of the JP black hole is often  approximated by considering only the leading non--vanishing contribution to $h(r)$ i.e. 
\begin{equation}
h(r) \approx \epsilon_3\Big{(}\frac{m}{r}\Big{)}^3. \label{eq:truncated}
\end{equation}
Throughout this paper we will either use the full expression for $h(r)$ viz. 
\ref{eq:SSdeviationh} 
or the truncated version,
\ref{eq:truncated}.

The Paczy{\'n}ski-Wiita potential for a black hole of mass, $m$, is \begin{equation}\label{eq:PWpotential}
	\Phi_{PW} := -\frac{m}{r-2m}.
\end{equation}
This potential has been used to successfully approximate the dynamics about a Schwarzschild black hole. One can obtain a pseudo--Newtonian potential for \emph{any} static spherically symmetric spacetime by the following argument \cite{faraoni2016paczynski}. Consider the spacetime
\begin{equation}
	ds^2 = g_{00}(r)dt^2 + g_{11}(r)dr^2 + r^2 d\theta^2+r^2 \sin^2\theta d\phi^2.
\end{equation}
By examining the orbits of a test particle in this spacetime one can identify an effective gravitational potential of the form
\begin{equation}
	\Phi := \frac{1}{2}(g^{00}(r) + k),
\end{equation}
where $k\in\mathbb{R}$ is an arbitrary constant. When applied to the Schwarzschild spacetime, the choice $k=1$ yields the original Paczy{\'n}ski-Wiita potential.

Turning our attention to the spherically-symmetric Johannsen-Psaltis metric Eq. \ref{eq:SSJPmetric}, we find the corresponding potential when again choosing $k=1$ is
\begin{equation}
	\Phi_{JP} := -\frac{m}{r-2m} + \frac12\frac{rh(r)}{[1 + h(r)](r-2m)}. \label{phiJP}
\end{equation}
This clearly reduces to $\Phi_{PW}$ when $h(r)=0$.

\section{Accretion onto a static Johannsen--Psaltis black hole}

We model the gas as a perfect fluid with a polytrope equation of state, $p = K \rho^{\gamma}$, where the adiabatic index, $\gamma$, is given by the ratio of specific heats i.e. $\gamma = c_{p} / c_{v}$ The dynamical motion of the fluid is determined by the conservation of mass and momentum. In spherical symmetry, and considering only radially infalling steady-state flows we obtain
\begin{eqnarray}
	\frac{1}{r^2}\frac{d}{dr}(\rho u r^2) = 0,\label{eq:continuity} \\
	u\frac{du}{dr} + \frac{1}{\rho}\frac{dp}{dr}+\frac{d\Phi_{JP}}{dr} = 0,\label{eq:mattermotion} 
\end{eqnarray}
where $u(r)$ is the radial velocity of the fluid. We now rewrite the equation of state using the polytropic index, $p = K \rho^{1 + \frac{1}{n}}$.
Integrating Eqs. \ref{eq:continuity} and \ref{eq:mattermotion} we find the conserved mass flux and specific energy respectively to be
\begin{eqnarray}
	\dot{M}     = \rho u r^2,\label{eq:Mdot}	
\\
	\mathcal{E} = \frac12u^2 + na^2 + \Phi_{JP} = na_\infty^2, \label{eq:energy}
\end{eqnarray}
where we have denoted $\displaystyle a=\sqrt{\frac{d p}{d \rho}}$ as the adiabatic sound speed \cite{chakrabarti1990theory,shapiro2008black}. 
Our boundary conditions are determined at a point, $r_{\infty}$, which is sufficiently far away from the black hole that the fluid velocity ($u(r_{\infty}) \equiv u_{\infty}$) and the gravitational potential both vanish. The fluid sound speed ``at infinity'' is denoted by $a_\infty$.
 Further, we rewrite \ref{eq:Mdot} to obtain the accretion rate
\begin{equation}\label{eq:accretion}
	\dot\mathcal{M} = a^{2n}ur^2,
\end{equation}
which is related to the mass flux via $\dot\mathcal{M} = \dot M\gamma^nK^n$ \cite{chakrabarti1990theory}. 
We now combine Eqs. \ref{eq:energy} and \ref{eq:accretion} to obtain
\begin{equation}
	\frac{du}{dr} = \frac{\displaystyle\frac{2a^2}{r} 
					- \frac{d\Phi_{JP}}{dr}}{\displaystyle u - \frac{a^2}{u}}. \label{wind}
\end{equation}

To ensure that the accretion flow is smooth, the denominator of \ref{wind} must vanish at the same point as the numerator. This \emph{critical point}, $r_{c}$ is determined by the conditions:
\begin{equation}\label{eq:criticalpconds}
	a_c = u_c,\qquad \frac{d\Phi_{JP}}{dr}\bigg |_{r=r_c} = \frac{2a_c^2}{r_c}.
\end{equation}
The first condition tells us that the critical point is also a \emph{sonic point}. If the fluid accelerates from rest  far away from the black hole ($u_{\infty}=0$) it must reach its local sound speed at the critical point i.e. $u_{c} = a_{c}$.

Assuming $h(r)$ is small, we can solve for the critical points by linearising \ref{eq:criticalpconds} in $h(r)$ to obtain the polynomial
\begin{equation}\label{eq:infinitepoly}
	2[m r_c-2a_c^2(r_c-2m)^2]+r_{c}h(r_c)(4m-3r_c)=0,
\end{equation}
and then compute the roots. This equation is valid for the full form of $h(r)$ given by \ref{eq:SSdeviationh}, but this clearly would be impossible to solve. Thus we only consider the leading contribution to $h(r)$ given by \ref{eq:truncated}, which is commonly invoked in studies of the JP spacetime \cite{johannsen2011metric,zelenka2017chaotic}.

The specialisation
\begin{equation}
	h_{c}(r_{c}) = \epsilon_3\Big{(}\frac{m}{r_{c}}\Big{)}^3.
\end{equation}
 reduces Eq. \ref{eq:infinitepoly} to a quartic polynomial in $\alpha := r_{c}/m$ which has the form
\begin{equation} \label{eq:quartic}
	-2\epsilon_3 + \frac{3}{2}\epsilon_3\alpha + 8a_c^2\alpha^2 - (1 + 8a_{c}^{2})\alpha^3 + 2a_c^2\alpha^4 = 0.
\end{equation}
 The four roots of this quartic are lengthy expressions and are hard to interpret intuitively. We observed that for a wide range of values for $\epsilon_3$ and $a_{c}$ only one of the roots is always real and outside the event horizon located at $r=2m$\footnote{The location of the event horizon in the static limit of the Johannsen-Psaltis metric is still at $r=2m$.}. We identify this root as \emph{the} critical point of the accretion flow.

To obtain a more tractable analytical expression for this root, we linearise \ref{eq:quartic} about $\epsilon_3$ for the rest of this section. Further, taking our equations to be linear in $a_{c}^{2}$, as is done in \cite{shapiro2008black}, gives the expression
\begin{equation}\label{eq:criticalpointrfinal}
	r_c \approx m(4 + \frac{1}{2a_c^2} - \frac{\epsilon_3}{2}),
\end{equation}
which reduces to the well known result for the Schwarzschild spacetime found in the literature \cite{shapiro2008black, michel1972}.

Using Eq. \ref{eq:criticalpointrfinal} and Eq. \ref{eq:energy} and linearising about $\epsilon_3$, $a_{c}^{2}$ and $a^2$ we find the relationship between the critical sound speed and the boundary condition as
\begin{equation}
	a_c^2 = \frac{2n}{2n-3} a_{\infty},
\end{equation}
which is the same as the Schwarzschild case.

Using the above together with Eq. \ref{eq:accretion} we find the Bondi accretion rate,
\begin{equation}
	\dot{\mathcal{M}} = 2^{\frac12 + n}m^2\Big{(}\frac{na_\infty^2}{2n-3}\Big{)}^{\frac12 + n}
						\Big{(}4 + \frac{2n-3}{4na_\infty^2}-\frac{\epsilon_3}{2}\Big{)}^2.
\end{equation}

\section{Results}

In the absence of scalar hair i.e. $h(r) = 0$, we recover the accretion rate for a fluid accelerated by a Paczy{\'n}ski--Wiita potential \cite{chakrabarti1990theory}.
The scalar hair parameter $\epsilon_3$ influences the accretion rate even when small. When $\epsilon_3>0$ the accretion rate is lower than that for a Paczy{\'n}ski--Wiita potential, while when $\epsilon_3<0$ the accretion rate is higher. This result was anticipated by a simple back--of--the--envelope calculation of the gravitational force of our system. For the potential, $\Phi_{JP}$, we have $F \sim - \Phi_{JP}' \sim -\frac{1}{r^{2}} + \epsilon_{3} \frac{1}{r^{4}}$. For positive values of $\epsilon_{3}$ the gravitational force becomes less negative and hence less attractive. Similarly, negative values of $\epsilon_{3}$ result in a more negative and thus larger attractive force.

When one analyses geodesic motion in the spacetime \ref{eq:SSJPmetric} one finds that the trajectories of test particles is consistent with our accretion results. When $\epsilon_{3}>0$ radially infalling particles experience lower accelerations than they do in the Schwarzschild spacetime. Similarly when $\epsilon_{3}<0$ test particles in the JP spacetime fall towards the black hole at a faster rate than they do in a Schwarzschild geometry. This analysis is contained in our more detailed companion paper \cite{john2019bondi}.

\section{Summary and ties to experiment}

We examined the problem of accretion onto a static black hole that generically violates the no hair theorem. The Johannsen--Psaltis(JP) metric admits scalar hair which parametrises deviations from the Kerr metric. Crucially the JP metric is generic in the sense that it does not arise from a specific theory of the gravity.

We considered the static limit of the JP metric and solved a Newtonian version of the accretion problem for this black hole. The Paczy{\'n}ski--Wiita potential has been used to successfully approximate dynamics around a Schwarzschild black hole. A recently devised recipe \cite{faraoni2016paczynski} generalised the Paczy{\'n}ski--Wiita potential to arbitrary, static, spherically symmetric black hole spacetimes. We applied this technique to obtain a pseudo--Newtonian potential, $\Phi_{JP}$, that mimics the gravitational behaviour of a static JP black hole.

We modelled accretion onto a static JP black hole by considering a fluid accelerated radially inwards by the potential, $\Phi_{JP}$. Initially at rest, the fluid reaches its local sound speed at a critical point and then proceeds to accelerate towards the event horizon at supersonic speeds. We obtained an approximate analytical expression for the location of the critical point by solving a fourth order polynomial equation. Only one of the four roots was found to be real, positive and outside the event horizon. We identified this root as \emph{the} critical point of the flow. 

We obtained an analytical expression for the accretion rate, $\dot{\mathcal{M}}$ given by \ref{eq:accretion}. The accretion rate is proportional to the square of the black hole mass i.e. $\dot{\mathcal{M}} \sim m^{2}$ which is a common feature in Bondi accretion. Positive(negative) values of the scalar hair parameter, $\epsilon_{3}$, were found to reduce(increase) the accretion rate. Our results can be used to vindicate the validity of the pseudo--Newtonian approach to black hole physics. 

The problem we investigated was an idealised study of accretion onto a JP black hole. This framework can be used to incorporate various important physical phenomena like radiative processes, viscous dissipation, magnetic fields and accretion disks.

For example, in gravitational lensing induced by a black hole, a so--called photon ring is predicted at $r \approx 10 M$. This arises from null geodesics wrapping around the black hole multiple times and is predicted to be prominent and highly circular and fairly independent of the microphysics of the accretion flow. Even the modest presence of scalar hair (for example a non--zero $\epsilon_{3}$) will induce appreciable eccentricity in the photon ring \cite{johannsen2013photon}. The Event Horizon Telescope has obtained an upper limit \cite{akiyama2019firstI} on the fractional quadrupole moment deviation, $\Delta Q/ Q \lesssim 4$. It is anticipated that these constraints will improve with greater resolution and millimeter wavelength very long baseline interferometry.

As another application, fluorescent iron lines occur in the spectra of gas accreting onto a rotating black hole. This spectrum has a characteristic shape that is altered when the black hole is deformed. Measuring deviations in the form of quadrupole deformations of the metric to a precision of a factor of unity will require measuring the spectra to a precision of about 5\%. This is achievable with planned X--ray observatories Astro--H and ATHENA+ \cite{johannsen2013testing}. 

\bibliography{refs2}

\providecommand{\href}[2]{#2}\begingroup\raggedright\begin{thebibliography}{10}

\bibitem{Will2014}
C.~M. Will, \emph{The confrontation between general relativity and experiment},
  \href{https://doi.org/10.12942/lrr-2014-4}{\emph{Living Reviews in
  Relativity} {\bfseries 17} (2014) 4}.

\bibitem{abbott2016observation}
B.~P. Abbott, R.~Abbott, T.~Abbott, M.~Abernathy, F.~Acernese, K.~Ackley
  et~al., \emph{Observation of gravitational waves from a binary black hole
  merger}, {\emph{Phys. {R}ev. {L}ett.} {\bfseries 116} (2016) 061102}.

\bibitem{nohair1}
W.~Israel, \emph{Event horizons in static vacuum space-times},
  \href{https://doi.org/10.1103/PhysRev.164.1776}{\emph{Phys. Rev.} {\bfseries
  164} (1967) 1776}.

\bibitem{nohair2}
W.~Israel, \emph{Event horizons in static electrovac space-times},
  \href{https://doi.org/10.1007/BF01645859}{\emph{Communications in
  Mathematical Physics} {\bfseries 8} (1968) 245}.

\bibitem{nohair3}
B.~Carter, \emph{Axisymmetric black hole has only two degrees of freedom},
  \href{https://doi.org/10.1103/PhysRevLett.26.331}{\emph{Phys. Rev. Lett.}
  {\bfseries 26} (1971) 331}.

\bibitem{berti2015testing}
E.~Berti, E.~Barausse, V.~Cardoso, L.~Gualtieri, P.~Pani, U.~Sperhake et~al.,
  \emph{Testing general relativity with present and future astrophysical
  observations}, {\emph{Classical and Quantum Gravity} {\bfseries 32} (2015)
  243001}.

\bibitem{broderick2014testing}
A.~E. Broderick, T.~Johannsen, A.~Loeb and D.~Psaltis, \emph{Testing the
  no-hair theorem with event horizon telescope observations of {S}agittarius
  {A}}, {\emph{The Astrophysical Journal} {\bfseries 784} (2014) 7}.

\bibitem{cunha2015shadows}
P.~V. Cunha, C.~A. Herdeiro, E.~Radu and H.~F. R{\'u}narsson, \emph{{Shadows of
  Kerr black holes with scalar hair}}, {\emph{Physical Review Letters}
  {\bfseries 115} (2015) 211102}.

\bibitem{akiyama2019firstI}
K.~Akiyama, A.~Alberdi, W.~Alef, K.~Asada, R.~Azulay, A.-K. Baczko et~al.,
  \emph{{First M87 Event Horizon Telescope Results. I. The Shadow of the
  Supermassive Black Hole}}, {\emph{The Astrophysical Journal Letters}
  {\bfseries 875} (2019) L1}.

\bibitem{akiyama2019firstII}
K.~Akiyama, A.~Alberdi, W.~Alef, K.~Asada, R.~Azulay, A.-K. Baczko et~al.,
  \emph{{First M87 Event Horizon Telescope Results. II. Array and
  Instrumentation}}, {\emph{The Astrophysical Journal Letters} {\bfseries 875}
  (2019) L2}.

\bibitem{akiyama2019firstIII}
K.~Akiyama, A.~Alberdi, W.~Alef, K.~Asada, R.~Azulay, A.-K. Baczko et~al.,
  \emph{{First M87 Event Horizon Telescope Results. III. Data Processing and
  Calibration}}, {\emph{The Astrophysical Journal Letters} {\bfseries 875}
  (2019) L3}.

\bibitem{akiyama2019firstIV}
K.~Akiyama, A.~Alberdi, W.~Alef, K.~Asada, R.~Azulay, A.-K. Baczko et~al.,
  \emph{{First M87 Event Horizon Telescope Results. IV. Imaging the Central
  Supermassive Black Hole}}, {\emph{The Astrophysical Journal Letters}
  {\bfseries 875} (2019) L4}.

\bibitem{akiyama2019firstV}
K.~Akiyama, A.~Alberdi, W.~Alef, K.~Asada, R.~Azulay, A.-K. Baczko et~al.,
  \emph{{First M87 Event Horizon Telescope Results. V. Physical Origin of the
  Asymmetric Ring}}, {\emph{The Astrophysical Journal Letters} {\bfseries 875}
  (2019) L5}.

\bibitem{akiyama2019firstVI}
K.~Akiyama, A.~Alberdi, W.~Alef, K.~Asada, R.~Azulay, A.-K. Baczko et~al.,
  \emph{{First M87 Event Horizon Telescope Results. VI. The Shadow and Mass of
  the Central Black Hole}}, {\emph{The Astrophysical Journal Letters}
  {\bfseries 875} (2019) L6}.

\bibitem{moffat2006scalar}
J.~W. Moffat, \emph{Scalar--tensor--vector gravity theory}, {\emph{Journal of
  Cosmology and Astroparticle Physics} {\bfseries 2006} (2006) 004}.

\bibitem{teves}
J.~D. Bekenstein, \emph{Relativistic gravitation theory for the modified
  {N}ewtonian dynamics paradigm},
  \href{https://doi.org/10.1103/PhysRevD.70.083509}{\emph{Phys. Rev. D}
  {\bfseries 70} (2004) 083509}.

\bibitem{nojiri}
S.~Nojiri and S.~D. Odintsov, \emph{{Modified f(R) gravity consistent with
  realistic cosmology: From a matter dominated epoch to a dark energy
  universe}}, {\emph{Physical Review D} {\bfseries 74} (2006) 086005}.

\bibitem{johannsen2011metric}
T.~Johannsen and D.~Psaltis, \emph{Metric for rapidly spinning black holes
  suitable for strong-field tests of the no-hair theorem}, {\emph{Physical
  Review D} {\bfseries 83} (2011) 124015}.

\bibitem{manko1992generalizations}
V.~Manko and I.~D. Novikov, \emph{Generalizations of the {K}err and
  {K}err-{N}ewman metrics possessing an arbitrary set of mass-multipole
  moments}, {\emph{Classical and Quantum Gravity} {\bfseries 9} (1992) 2477}.

\bibitem{paczynsky1980}
B.~Paczynsky and P.~J. Wiita, \emph{Thick accretion disks and supercritical
  luminosities}, {\emph{Astronomy and Astrophysics} {\bfseries 88} (1980) 23}.

\bibitem{abramowicz2009paczynski}
M.~A. Abramowicz, \emph{The {P}aczy{\'n}ski-{W}iita potential. {A} step-by-step
  {``}derivation{''}-{C}ommentary on: {P}aczy{\'n}ski {B}. and {W}iita {PJ},
  1980, {A}\&{A}, 88, 23}, {\emph{Astronomy \& Astrophysics} {\bfseries 500}
  (2009) 213}.

\bibitem{faraoni2016paczynski}
V.~Faraoni, S.~D. Belknap-Keet and M.~Lapierre-Leonard,
  \emph{Paczynski-{W}iita-like potential for any static spherical black hole in
  metric theories of gravity}, {\emph{Physical Review D} {\bfseries 93} (2016)
  044042}.

\bibitem{tejeda2014}
E.~Tejeda and S.~Rosswog, \emph{Generalized {N}ewtonian description of particle
  motion in spherically symmetric spacetimes}, {\emph{arXiv preprint
  arXiv:1402.1171} (2014) }.

\bibitem{chakrabarti1990theory}
S.~K. Chakrabarti, \emph{Theory of transonic astrophysical flows}. World
  Scientific, 1990.

\bibitem{shapiro2008black}
S.~L. Shapiro and S.~A. Teukolsky, \emph{{Black holes, white dwarfs, and
  neutron stars: The physics of compact objects}}. John Wiley \& Sons, 2008.

\bibitem{zelenka2017chaotic}
O.~Zelenka and G.~Loukes-Gerakopoulos, \emph{Chaotic motion in the
  {J}ohannsen-{P}saltis spacetime}, {\emph{arXiv preprint arXiv:1711.02442}
  (2017) }.

\bibitem{michel1972}
F.~C. Michel, \emph{Accretion of matter by condensed objects},
  {\emph{Astrophysics and Space Science} {\bfseries 15} (1972) 153}.

\bibitem{john2019bondi}
A.~John and C.~Stevens, \emph{{Bondi accretion in the spherically symmetric
  Johannsen-Psaltis spacetime}}, {\emph{arXiv preprint arXiv:1903.01958} (2019)
  }.

\bibitem{johannsen2013photon}
T.~Johannsen, \emph{{Photon rings around Kerr and Kerr-like black holes}},
  {\emph{The Astrophysical Journal} {\bfseries 777} (2013) 170}.

\bibitem{johannsen2013testing}
T.~Johannsen and D.~Psaltis, \emph{{Testing the no-hair theorem with
  observations in the electromagnetic spectrum. IV. Relativistically broadened
  iron lines}}, {\emph{The Astrophysical Journal} {\bfseries 773} (2013) 57}.

\end{thebibliography}\endgroup

\end{document}